\documentclass[aps,prl,reprint,superscriptaddress,amsmath,amssymb,longbibliography]{revtex4-1}
\usepackage{graphicx}
\usepackage{braket} 
\usepackage[colorlinks=true,urlcolor=blue,citecolor=blue,linkcolor=blue,bookmarks=false,pdfstartview={FitH}]{hyperref}
\usepackage{capt-of}
\usepackage{bm}
\usepackage{bbold}

\begin{document}
	\title{Analogy between the ``Hidden Order'' and the Orbital Antiferromagnetism in URu$_{2-x}$Fe$_x$Si$_2$}
	\author{H.-H.~Kung}
	\email{hk458@physics.rutgers.edu}
	\affiliation{Department of Physics \& Astronomy, Rutgers University, Piscataway, New Jersey 08854, USA}
	\author{S.~Ran}
	\author{N.~Kanchanavatee}
	\affiliation{Department of Physics, University of California San Diego, La Jolla, California 92093, USA}
	\affiliation{Center for Advanced Nanoscience, University of California San Diego, La Jolla, California 92093, USA}
	\author{V.~Krapivin}
	\author{A.~Lee}
	\affiliation{Department of Physics \& Astronomy, Rutgers University, Piscataway, New Jersey 08854, USA}
	\author{J.~A.~Mydosh}
	\affiliation{Kamerlingh Onnes Laboratory, Leiden University, 2300 RA Leiden, The Netherlands}
	\author{K.~Haule}
	\affiliation{Department of Physics \& Astronomy, Rutgers University, Piscataway, New Jersey 08854, USA}
	\author{M.~B.~Maple}
	\affiliation{Department of Physics, University of California San Diego, La Jolla, California 92093, USA}
	\affiliation{Center for Advanced Nanoscience, University of California San Diego, La Jolla, California 92093, USA}
	\author{G.~Blumberg}
	\email{girsh@physics.rutgers.edu}
	\affiliation{Department of Physics \& Astronomy, Rutgers University, Piscataway, New Jersey 08854, USA}
	\affiliation{National Institute of Chemical Physics and Biophysics, 12618 Tallinn, Estonia}
	%
	\begin{abstract}
		We study URu$_{2-x}$Fe$_x$Si$_2$, in which two types of staggered phases compete at low temperature as the iron concentration $x$ is varied:
		the nonmagnetic ``hidden order'' (HO) phase below the critical concentration $x_c$, and unconventional antiferromagnetic (AF) phase above $x_c$.
		By using polarization resolved Raman spectroscopy, we detect a collective mode of pseudovector-like $A_{2g}$ symmetry whose energy continuously evolves with increasing $x$; 
		it monotonically decreases in the HO phase until it vanishes at $x=x_c$, and then reappears with increasing energy in the AF phase. 
		The mode's evolution provides direct evidence for unified order parameter for both nonmagnetic and magnetic phases arising from the orbital degrees-of-freedom of the uranium-5$f$ electrons.
	\end{abstract}
	\maketitle
	%

URu$_2$Si$_2$ holds long-standing interest in the strongly correlated electron community due to several emergent types of long range orders it exhibits. 
Below the second order phase transition temperature $T_\text{DW}(x)$, two density-wave-like phases involving long range ordering of the uranium-5$f$ electrons compete when a critical parameter $x$ is tuned~\cite{Hall2015}, where $x$ can be chemical substituent concentration~\cite{Kanchanavatee2011,Ran2016}, pressure~\cite{Butch2010,Bourdarot2011} or magnetic field~\cite{Jaime2002,Aoki2009}.
At $x<x_c$, the system settles in the enigmatic ``hidden order'' (HO) phase~\cite{Palstra1985,Maple1986,Schlabitz1986}, which transforms into an unconventional large moment antiferromagnetic (LMAF) phase through a first order transition for $x>x_c$.
Below 1.5\,K, a superconducting state, which likely breaks time reversal symmetry~\cite{Schemm2015}, emerges from the HO phase.

Recently, much effort has been dedicated towards unraveling the order parameter of the HO phase through several newly developed experimental and theoretical techniques~\cite{Aynajian2010,Schmidt2010,Okazaki2011,Riggs2015,Schemm2015,Mydosh2011}.
In particular, the symmetry analysis of the low temperature Raman scattering data implies that the reflection symmetries of tetragonal $D_{4h}$ point group (No.~139 $I4/mmm$) associated with the paramagnetic (PM) state are broken, and that a chirality density wave emerges as the HO ground state~\cite{Kung2015}.
\begin{figure*}
	\includegraphics[width=16cm]{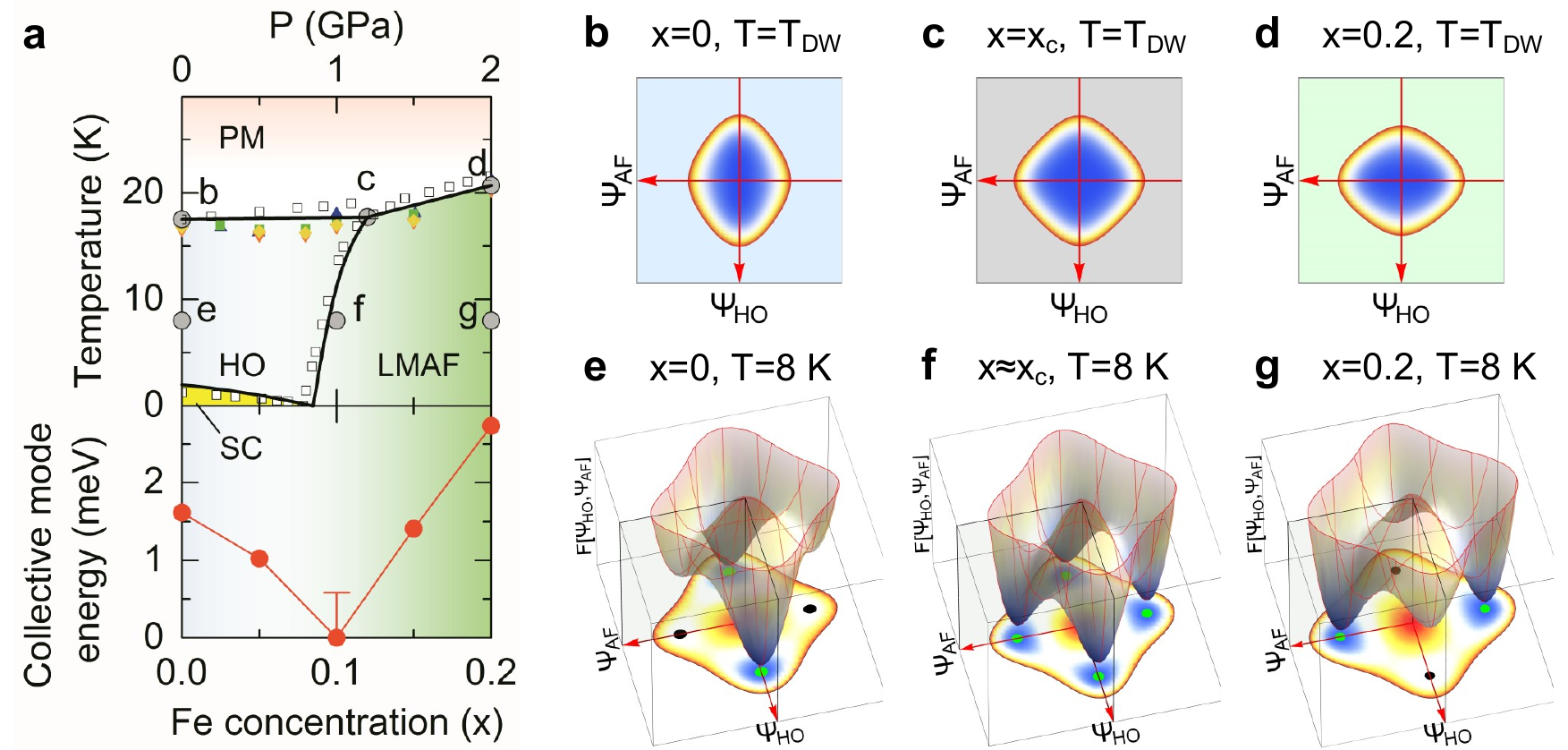}
	\caption{\label{fig:1}
		(a)~The upper panel shows the phase diagram of URu$_{2}$Si$_2$ system, where the black lines show the phase boundaries.
		The measurements on the iron substituted URu$_{2-x}$Fe$_x$Si$_2$ crystals from neutron diffraction~\cite{Das2015} (blue triangle), electrical resistivity~\cite{Kanchanavatee2011} (green square), magnetic susceptibility~\cite{Kanchanavatee2011} (purple triangle), and heat capacity~\cite{Ran2016} (yellow diamond), are overlaid with the neutron diffraction results for URu$_{2}$Si$_2$ under hydrostatic pressure~\cite{Butch2010} (open square) to show the similarity between the two tuning parameters.
		The lower panel shows the dependence of the $A_{2g}$ collective mode energy on the Fe concentration, $x$ [Fig.~\ref{fig:2}].
		At the critical concentration, $x=0.1$, the mode maximum is below the accessible energy cutoff.
		Therefore, the data point is placed at zero energy, with the error bar reflecting the instrumental cutoff.
		(b)--(g)~Schematics of the Ginzburg-Landau free energy in Eq.~(\ref{eq:1}) at various special points in the phase diagram [solid gray circles in (a)].
		$\psi_\text{HO}$ and $\psi_\text{AF}$ are the real and imaginary part of the hexadecapole order parameter, respectively~\cite{Haule2009,Haule2010}.
	}
\end{figure*}

The HO and LMAF phases are known to exhibit ``adiabatic continuity''~\cite{Jo2007}; i.e., both phases possess similar electronic properties~\cite{Hassinger2008,Kanchanavatee2011}, and the Fermi surface practically shows no change across the phase boundary~\cite{Jo2007}.
Furthermore, inelastic neutron scattering observed a dispersive collective excitation in the HO phase~\cite{Broholm1987,Bourdarot2011} and recently in the LMAF phase of pressurized URu$_2$Si$_2$~\cite{Williams2016A}.
This raises the intriguing question of the symmetry relation between the two phases.
However, experimental progress is hindered due to inherent constraints of low temperature pressurized experiments.

The availability of URu$_{2-x}$Fe$_x$Si$_2$ crystals~\cite{Kanchanavatee2011,Ran2016} made it possible to perform high-resolution spectroscopic experiments at low temperature and ambient pressure in both the HO and LMAF phases.
Iron substitution mimics the effect of applying small pressure or in-plane stress on the URu$_2$Si$_2$ lattice, and the iron (Fe) concentration, $x$, can be approximately treated as an effective ``chemical pressure''~\cite{Kanchanavatee2011}.
Recently, the phase digram of URu$_{2-x}$Fe$_x$Si$_2$ single crystals have been determined~\cite{Wilson2015,Hall2015,Das2015,Ran2016,Butch2016}, which resembles the low pressure phase diagram of pristine URu$_2$Si$_2$~\cite{Butch2010,Mydosh2011} [Fig.~\ref{fig:1}(a)].
The inelastic neutron scattering measurements again illustrate the analogies of the LMAF phase to the HO phase~\cite{Butch2016,Williams2016B}, albeit differences remain relating to the existence of the resonance in the LMAF state of pressurized~\cite{Williams2016A,Williams2016B} or Fe-substituted crystals~\cite{Butch2016}.

In this Letter, we study the dynamical fluctuations between the competing nonmagnetic HO and the time-reversal-symmetry breaking LMAF ground states in URu$_{2-x}$Fe$_x$Si$_2$ as a function of $x$ using polarization resolved Raman spectroscopy~\cite{SM}.
Albeit the distinct discrete symmetries are broken above and below the critical concentration $x_c$, we detect a collective mode continuously evolving with parameter $x$ in the pseudovector-like $A_{2g}$ symmetry channel.
In the HO phase, the mode energy decreases as $x$ is increased, disappearing at the critical Fe concentration $x_c$.
In the LMAF phase, the collective mode again emerges in the same $A_{2g}$ symmetry channel with the energy increasing with $x$.
The continues transformation of this collective excitation, a photoinduced transition between the HO and LMAF electronic phases, provides direct experimental evidence for a unified order parameter for both nonmagnetic and magnetic phases arising from the orbital degree of freedom of the uranium-5$f$ electrons.

The polarized Raman spectra were acquired in a quasibackscattering geometry from the \textit{ab} surface of URu$_{2-x}$Fe$_x$Si$_2$ single crystals grown by the Czochralski method~\cite{SM}.
We use 752.5\,nm line of a Kr$^{+}$ laser for excitation. 
The scattered light was analyzed by a custom triple-grating spectrometer.
The laser spot size on the sample is roughly $50\times 100\,\mu\text{m}^2$.
The power on the sample is about 12\,mW for most temperatures, and kept below 6\,mW to achieve the lowest temperatures.

\begin{figure*}[t]
	\includegraphics[width=16cm]{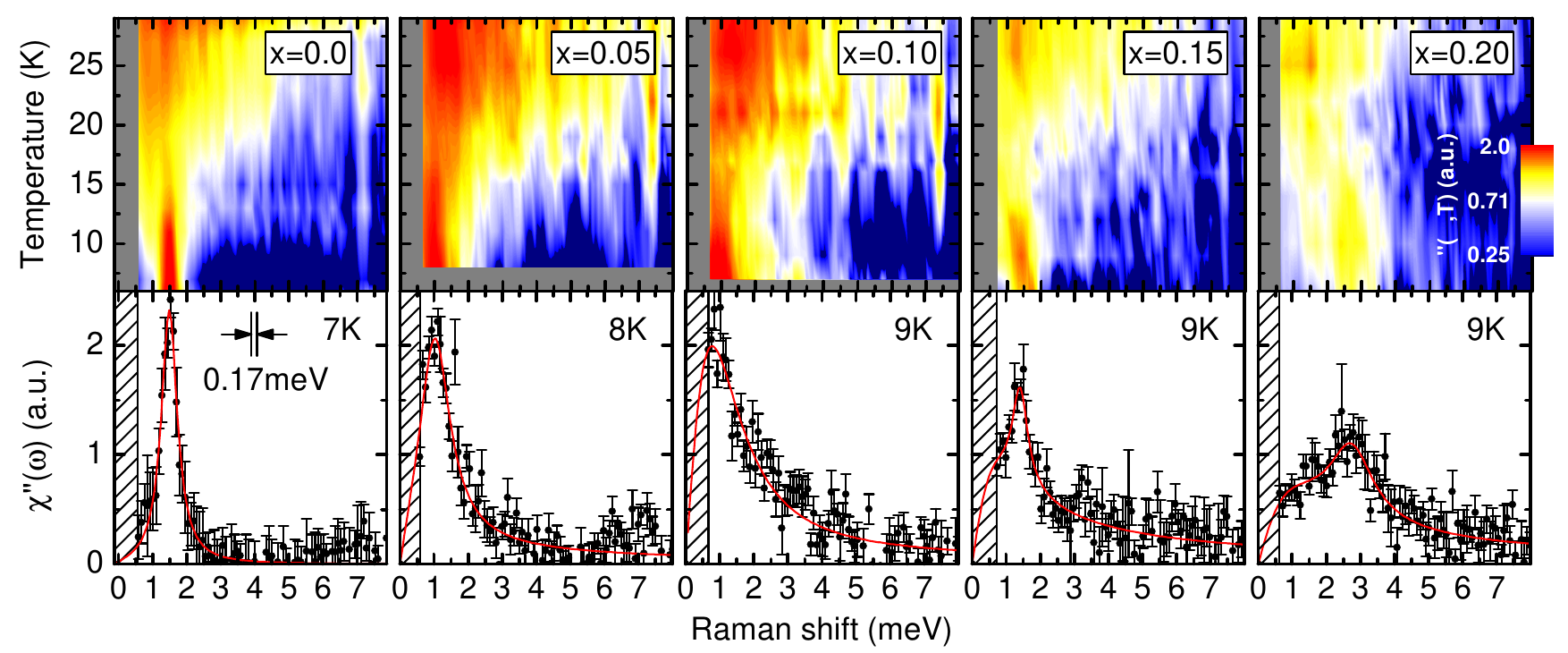}
	\caption{\label{fig:2}
		Low temperature Raman response in the $A_{2g}$ symmetry channel, $\chi^{\prime\prime}_{A2g}(\omega,T)$~\cite{SM}.
		The upper panels show intensity plots, where the intensities are color coded in logarithmic scale.
		The lower panels show the spectra at about half the transition temperature to emphasize the collective mode, where the error bars represent one standard deviation, and the red solid lines are guides to the eye.
		The energies of this mode as function of the Fe concentration $x$ are shown in Fig.~\ref{fig:1}(a).}
\end{figure*}

Figure~\ref{fig:2} shows the temperature dependence of the Raman response in the eminent $A_{2g}$ symmetry channel of the $D_{4h}$ group, which transforms as a pseudovector~\cite{Khveshchenko1994}.
The upper panels show the intensity plots of the low energy Raman response $\chi_{A2g}^{\prime\prime}(\omega,T)$ below 30\,K.
Above $T_\text{DW}(x)$, a quasielastic peak (QEP) comprises most of the spectral weight for all samples, narrowing towards the transition.
The observed QEP originates from overdamped excitations between quasidegenerate crystal field states~\cite{Kung2015,Haule2009}, and the narrowing of the QEP with cooling is due to the increase of excitation lifetime, related to the development of a hybridization gap and formation of a heavy Fermi liquid~\cite{Guo2012,Lobo2015}.

Below $T_\text{DW}(x)$, the most significant feature in the $A_{2g}$ channel is a sharp collective mode.
The sharpness of this resonance suggests the lack of relaxation channels due to the opening of an energy gap~\cite{Guo2012,Hall2012,Hall2015}.
In order to see the mode's lineshape more clearly, we plot $\chi_{A2g}^{\prime\prime}(\omega,T)$ for each Fe concentration $x$ in the lower panels, with $T\approx T_\text{DW}(x)/2$.
The lineshapes broaden with increasing $x$ owing to the inhomogeneity of the local stress field, or unsuppressed relaxation channels introduced by doping that interact with the collective mode, which may also be related to the increasing continuum in the $x=0.15$ and $0.2$ spectra.
In contrast to the monotonic broadening of the lineshape width, the collective mode frequency shows nonmonotonic behavior as function of $x$.
The mode energy versus Fe concentration $x$ is shown in the lower panel of Fig.~\ref{fig:1}(a).
The energy decreases with increasing $x$ in the HO phase, until it vanishes below the instrumental resolution at $x=0.10$, which is close to the HO and LMAF phase boundary determined by elastic neutron scattering~\cite{Das2015} and thermal expansion measurements~\cite{Ran2016}.
The resonance reappears in the LMAF phase, where the energy increases with increasing $x$.
The resonance in the LMAF state appears in the same $A_{2g}$ symmetry channel as the collective mode in the HO phase.

\begin{figure}[t]
	\includegraphics[width=7cm]{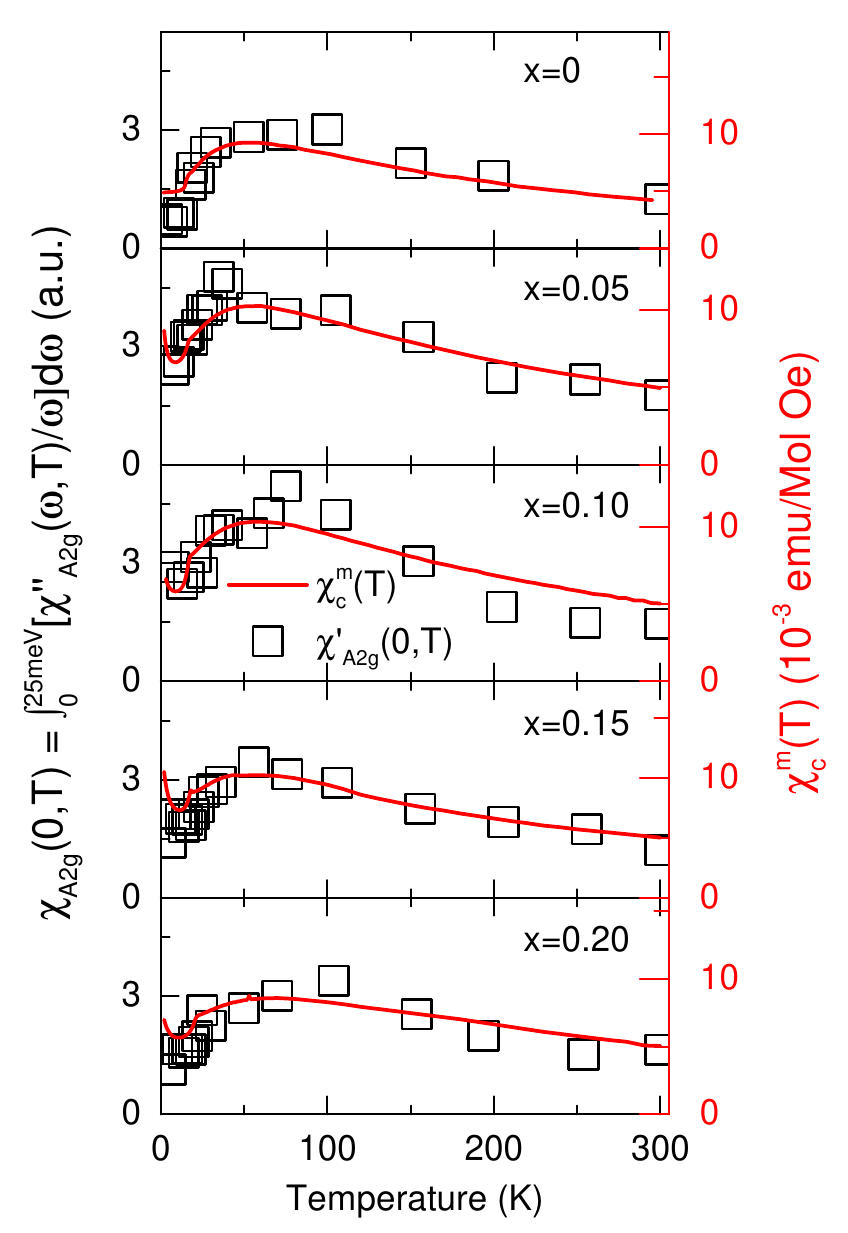}
	\caption{\label{fig:3}
		The static Raman susceptibility in the $A_{2g}$ symmetry channel (open squares) $\chi_{A2g}(0,T)$, 
		compared with the magnetic susceptibility with field applied along the \textit{c} axis~\cite{Ran2016} (solid line).
	}
\end{figure}

The similarity of the Raman response in the HO and LMAF phases encourages us to compare our results with the magnetic susceptibility.
Figure~\ref{fig:3} shows the temperature dependence of the real part of the static $A_{2g}$ Raman susceptibility $\chi_{A2g}(0,T)$, compared with the \textit{c} axis magnetic susceptibility $\chi_c^m(T)$~\cite{Ran2016}.
While there are discrepancies around the maxima at about 50--100\,K, both quantities follow the same Curie-Weiss-like temperature dependence above 100\,K, followed by a suppression approaching the second order phase transition.

The comparison between $\chi_{A2g}(0,T)$ and $\chi_c^m(T)$ has been studied within the frame work of a phenomenological minimal model~\cite{Haule2009,Kung2015}. 
The model is composed of two low-laying singlet orbital levels on uranium sites as suggested by recent experiment~\cite{Sundermann2016}, separated by small energy $\omega_0$.
These states with pseudovector-like $A_{2g}$ and full-symmetric $A_{1g}$ symmetries are denoted by $\ket{A_{2g}}$ and $\ket{A_{1g}}$, respectively.
At high temperatures, the crystal field states are quasidegenerate in energy and localized at the uranium $f$ shells in space.
The Curie-Weiss-like behavior above 100\,K in static magnetic-~\cite{Pfleiderer2006,Ran2016} and Raman-susceptibilities~\cite{Cooper1987,Buhot2014,Kung2015} suggest $A_{2g}$ pseudovector-like instabilities at low temperature.
Below about 50\,K, the Kondo screening begins setting in~\cite{Pfleiderer2006,Levallois2011,Guo2012,Hall2012,Mydosh2011} and the correlation length of the HO~\cite{Niklowitz2015} or LMAF~\cite{Butch2010,Niklowitz2010} phase builds at the ordering vector $Q_0=(0,0,1)$; therefore both the magnetic and Raman uniform susceptibilities start to decrease [Fig.~\ref{fig:3}].
Close to the transition temperature, both the HO and LMAF order parameters fluctuate regardless of the low temperature ordering [Figs.~\ref{fig:1}(b)--\ref{fig:1}(d)].
However, the static magnetic susceptibility at $Q_0$ diverges only across the PM--LMAF phase transition~\cite{Butch2010,Das2015}, whereas it becomes ``near critical'' from the PM-HO phase~\cite{Niklowitz2015}.
Thus, HO is a nonmagnetic transition, but there is the ``ghost'' of LMAF present as shown in Fig.~\ref{fig:1}(b).
Here, we find that the temperature dependencies of the static $A_{2g}$ Raman susceptibility $\chi_{A2g}(0,T)$ are similar and track $\chi_c^m(T)$ in all measured samples, suggesting that the minimal model is applicable for the studied Fe substituted crystals.

\begin{figure}[t]
	\includegraphics[width=8cm]{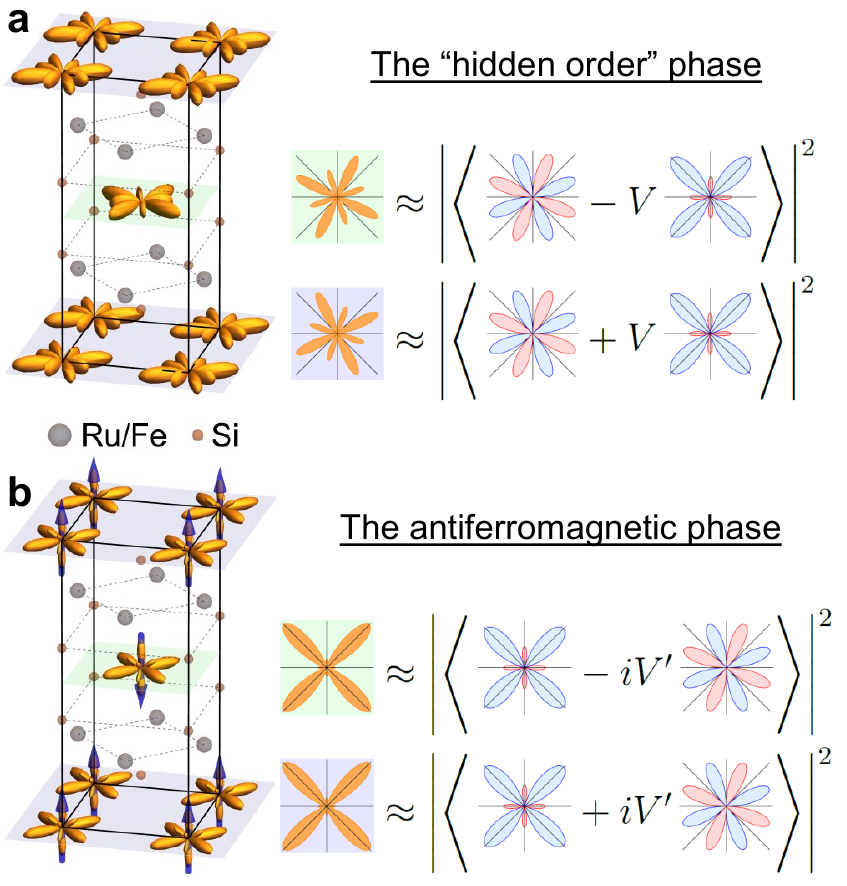}
	\caption{\label{fig:4}
		The crystal structure of URu$_{2-x}$Fe$_x$Si$_2$ in (a)~the HO and (b)~the LMAF phases.
		Illustrations capturing the symmetries of the charge distributions of the ground state wave functions are placed at the uranium atomic sites.
		On the right are illustrations showing the in-plane structures of the wave functions.
		In the HO phase, the crystal field state with the lowest energy has $A_{2g}$ symmetry with 8 nodal lines, $\ket{A_{2g}}$, which mixes with the first excited state with $A_{1g}$ symmetry, $\ket{A_{1g}}$, to form the local wave functions in the HO phase,
		$\ket{\text{HO}^\pm}\approx \cos\theta\ket{A_{2g}}\pm\sin\theta\ket{A_{1g}}$.
		In the LMAF phase, the ordering of the crystal field states switches, and the new wave functions in the LMAF phase are,
		$\ket{\text{AF}^\pm}\approx \cos\theta^\prime\ket{A_{1g}}\pm i\sin\theta^\prime\ket{A_{2g}}$.
		Here, $\theta\equiv\arcsin(V/\omega_0)$ and $\theta^\prime\equiv\arcsin(V^\prime/\omega_0)$, respectively. 
		$\omega_0$ is the splitting between the lowest lying crystal field states in the minimal model.
		$V$ and $V^\prime$ are the order parameter strength in the HO and LMAF phases, respectively.
	}
\end{figure}

We now discuss the origin and the observed doping dependence of the collective mode in the ordered phases within a phenomenological Ginzburg-Landau approach.
Within the minimal model, the two order parameters can be constructed from $\ket{A_{2g}}$ and $\ket{A_{1g}}$~\cite{Haule2009}.
The HO phase was explained as the state in which the two levels mix, resulting in a lower symmetry point group on the uranium site, which breaks all vertical and diagonal reflection symmetry planes, and thus acquires left and right handedness.~\cite{Haule2009,Kung2015} 
The staggering of left and right handed solutions on the lattice gives rise to the chirality density wave~\cite{Kung2015} [Fig.~\ref{fig:4}(a)].
In the HO phase, the staggered condensate can be approximated by a form $\ket{\psi_\text{HO}}=\displaystyle\prod_{r=A~site}\ket{\text{HO}^+_r}\times\prod_{r=B~site}\ket{\text{HO}^-_r}$. 
Note that $\ket{\text{HO}^{\pm}_r}$ at uranium site $r$ is dominantly $\ket{A_{2g}}$, with a small admixture of $\ket{A_{1g}}$, i.e., $\ket{\text{HO}^{\pm}} =\cos\theta\ket{A_{2g}} \pm \sin\theta\ket{A_{1g}}$.

In the HO phase the orbital mixing is purely real. If, however the mixing is purely imaginary, the charge distribution on the uranium site does not break any spatial symmetry; instead, it acquires nonzero out-of-plane magnetic moments, and thereby breaks time reversal symmetry.
The N\'{e}el-type condensate [Fig.~\ref{fig:4}(b)] takes the form
$\ket{\psi_\text{AF}}=\displaystyle\prod_{r=A~site}\ket{\text{AF}^+_r}\times\prod_{r=B~site}\ket{\text{AF}^-_r}$, where
$ \ket{\text{AF}^\pm}=\cos\theta^\prime\ket{A_{1g}}\pm i \sin\theta^\prime\ket{A_{2g}}$~\cite{Haule2009}. 
The two apparently competing orders, the chirality density wave and the antiferromagnetic state, are both constructed by mixing the two orbital wave functions on uranium sites with a real or an imaginary phase factor, $\sin\theta$ or $i\sin\theta^\prime$, thus unifying the two order parameters.

The Ginzburg-Landau free energy can then be constructed from the two component order parameter 
$\Psi^T\equiv \left(\begin{array}{cc}\psi_\text{HO}~&\psi_\text{AF}\end{array}\right)$, where the order parameters correspond to the two condensates $\ket{\psi_\text{HO}}$ and $\ket{\psi_\text{AF}}$ defined above.
The free energy takes the form
\begin{equation}
\label{eq:1}
F[\Psi]=\Psi^T \hat{A} \Psi + \beta \left( \Psi^T \Psi \right)^2
+ \gamma \left( \Psi^T \hat{\sigma}_1 \Psi \right)^2
\end{equation}
where $\hat{A}\equiv \begin{pmatrix}\alpha_{HO} & 0\\ 0 & \alpha_{AF} \end{pmatrix}$, with $\alpha_{HO}$ and $\alpha_{AF}$ vanish at the critical temperature.
$\hat{\sigma}_1\equiv \begin{pmatrix}0 & 1\\ 1 & 0 \end{pmatrix}$ is the Pauli matrix.
$\gamma$ controls a finite barrier between the two minima in Figs.~\ref{fig:1}(e)--\ref{fig:1}(g), and hence ensures phase separation between the HO and LMAF phases~\cite{Niklowitz2010}.
The free energy parameters are introduced following the recipes given in Haule and Kotliar~\cite{Haule2010,Boyer2016} with adjustments to match the phase diagram in Fig.~\ref{fig:1}(a)~\cite{SM}.

The Ginzburg-Landau free energy in the two dimensional space of $\psi_\text{HO}$ and $\psi_\text{AF}$ is shown in Figs.~\ref{fig:1}(b)--\ref{fig:1}(g).
Below the second order phase transition, two global and two local minima develop on $\psi_\text{HO}$ and $\psi_\text{AF}$ axes due to spontaneous discrete symmetry breaking, where the minima characterize the ground states in the HO and LMAF phases, respectively.

At the critical doping [Fig.~\ref{fig:1}(f)], the four minima are degenerate, but the barrier between the minima remains finite due to a $\gamma$ term in Ginzburg-Landau functional.
Therefore the transition between HO and LMAF phases is of the first order, and the coexistence of both phases is allowed, explaining the LMAF puddles that have been observed in the HO phase~\cite{Matsuda2001,Yokoyama2005}.

The energy separation between the dominant long range order (e.g., $\ket{\psi_\text{HO}}$)  and the sub-dominant order (e.g., $\ket{\psi_\text{AF}}$) is vanishingly small at the critical Fe concentration, and even away from this point can be smaller than the size of the gap. 
The exciton of subdominant symmetry  (e.g., $\ket{\psi_\text{AF}}$) can form in the gap, which then propagates through the order of the dominant symmetry (e.g., $\ket{\psi_\text{HO}}$). 
Likewise, when the ground state is of $\ket{\psi_\text{AF}}$, the propagating exciton is of $\ket{\psi_\text{HO}}$ symmetry.
The symmetry difference between the two condensates is $A_{2g}$; hence, such exciton can be detected by Raman in the $A_{2g}$ channel, and explains the sharp resonance shown in Fig.~\ref{fig:2}. 
It is clear from this discussion that the energy of the resonance vanishes at the critical Fe concentration, and is linearly increasing away from the critical point.
For superconductors, such an excitation is known as the Bardasis-Schrieffer mode, characterizing the transition between two competing Cooper pairing channels~\cite{Bardasis1961}.

More generally, the uranium-5$f$ orbitals in solids can arrange in surprising types of orders, including orders with broken chirality or time reversal symmetry.
While such orders are competing for the same phase space in URu$_2$Si$_2$, they are also subtly connected and were here unified into a common order parameter, which can be switched with small energy cost. 
The low energy excitations are usually Goldstone modes, but here we detected a new type of excitation, which connects two types of long range order, and is observed as a resonance by light scattering.
The resonance brings light to a long-standing problem of emergent phases of exotic local orbital self-organization and their interrelation.
\begin{acknowledgments}
We are grateful for discussions with C.~Broholm, N.P.~Butch, P.~Coleman, I.R.~Fisher, P.B.~Wiegmann and V.M.~Yakovenko.
G.B. and H.-H.K. acknowledge support from DOE BES Grant No. DE-SC0005463. 
A.L. and V.K. acknowledge NSF Grant No. DMR-1104884. 
K.H. acknowledges NSF Grant No. DMR-1405303.
M.B.M., S.R., and N.K. acknowledge DOE BES Grant No. DE-FG02-04ER46105 (crystal growth) and NSF Grant No. DMR-1206553 (materials characterization).
\end{acknowledgments}
\end{document}